\documentclass[11pt]{article}%
\usepackage{amsfonts}
\usepackage{amsmath}%
\usepackage{amssymb}%
\usepackage{graphicx}
\providecommand{\U}[1]{\protect\rule{.1in}{.1in}}

\begin{document}

\title{The Symplectic Camel and Quantum Universal Invariants: the Angel of Geometry
vs. the Demon of Algebra}
\author{Maurice A. de Gosson\thanks{maurice.de.gosson@gmail.com}\\University of Vienna, NuHAG}
\maketitle

\begin{abstract}
A positive definite symmetric matrix $\sigma$ qualifies as a quantum
mechanical covariance matrix if and only if $\sigma+\tfrac{1}{2}i\hbar
\Omega\geq0$ where $\Omega$ is the standard symplectic matrix. This well-known
condition is a strong version of the uncertainty principle, which can be
reinterpreted in terms of the topological notion of symplectic capacity,
closely related to Gromov's non-squeezing theorem. We show that a recent
refinement of the latter leads to a new class of geometric invariants. These
are the volumes of the orthogonal projections of the covariance ellipsoid on
symplectic subspaces of the phase space. We compare these geometric invariants
to the algebraic \textquotedblleft universal quantum
invariants\textquotedblright\ of Dodonov and Serafini.

\end{abstract}

\section{Introduction}

We consider a continuous variable system with $n$ degrees of freedom described
by a Hermitian positive operator $\widehat{\rho}$ on $L^{2}(\mathbb{R}^{n})$
with trace one. The phase space of the system is identified with
$\mathbb{R}^{2n}\equiv\mathbb{R}_{x}^{n}\times\mathbb{R}_{p}^{n}$ equipped
with the standard symplectic structure $z\wedge z^{\prime}=(z^{\prime}%
)^{T}\Omega z$ where $z=(x,p)$ and $\Omega=%
\begin{pmatrix}
0 & I\\
-I & 0
\end{pmatrix}
$. We will prove in this Letter the following geometric result: let
$\mathbb{F}_{2k}$ be an arbitrary symplectic subspace of $\mathbb{R}^{2n}$,
and denote by $\Pi_{\mathbb{F}_{2k}}$ the orthogonal projection operator of
$\mathbb{R}^{2n}$ onto $\mathbb{F}_{2k}$. Let $\sigma$ be a covariance matrix
satisfying the uncertainty problem in its strong form $\sigma+\tfrac{1}%
{2}i\hbar\Omega\geq0$ and let $W_{\sigma}:\frac{1}{2}z^{T}\sigma^{-1}z\leq1$
be the covariance ellipsoid of the system. Let us project this ellipsoid on
$\mathbb{F}_{2k}$. The volume of this projection satisfies
\[
\operatorname*{Vol}(\Pi_{\mathbb{F}_{2k}}W_{\sigma})\geq\frac{h^{k}}{2^{k}%
k!}.
\]
In particular, if $\mathbb{F}_{2k}$ is a plane of conjugate variables $x_{j}$,
$p_{j}$ (hence $k=1$) the projection is an ellipse of area not inferior to
$\frac{1}{2}h$, which is a geometric form of the uncertainty principle as we
have shown in previous work (\cite{physletta,Birk,FP}). In addition we compare
our results to Dodonov's and Serafini's \textquotedblleft quantum universal
invariants\textquotedblright\ \cite{dodo,serafini1,serafini2}, which comforts
us in our belief that the \textquotedblleft Angel of
Geometry\textquotedblright\ should always be preferred to the
\textquotedblleft Demon of Algebra\textquotedblright\ in conceptual questions.

\section{Strong Form of the Uncertainty Principle}

In what follows we will write the covariance matrix as
\begin{equation}
\sigma=%
\begin{pmatrix}
\sigma_{XX} & \sigma_{XP}\\
\sigma_{XP}^{T} & \sigma_{PP}%
\end{pmatrix}
\label{1}%
\end{equation}
where $\sigma_{XX}=(\sigma_{X_{j}X_{k}})_{1\leq j,k\leq n}$, $\sigma
_{XP}=\sigma_{PX}^{T}=(\sigma_{X_{j}P_{k}})_{1\leq j,k\leq n}$, $\sigma
_{PP}=(\sigma_{P_{j}P_{k}})_{1\leq j,k\leq n}$ are $n\times n$ matrices
($\sigma_{XX}$ and $\sigma_{PP}$ are symmetric).

A strong form of the uncertainty principle reads%
\begin{equation}
\sigma+\tfrac{1}{2}i\hbar\Omega\geq0 \label{2}%
\end{equation}
which is shorthand for saying that the Hermitian matrix $\sigma+\tfrac{1}%
{2}i\hbar\Omega$ is positive semidefinite. It is well-known that this
condition implies (but is not equivalent to) the Robertson--Schr\"{o}dinger
inequalities
\begin{equation}
\sigma_{X_{j}}^{2}\sigma_{P_{j}}^{2}\geq\sigma_{X_{j}P_{j}}^{2}+\tfrac{1}%
{4}\hbar^{2} \label{3}%
\end{equation}
where $\sigma_{X_{j}}^{2}=\sigma_{X_{j}X_{j}}$ and $\sigma_{P_{j}}^{2}%
=\sigma_{P_{j}P_{j}}$. It also implies that the covariance matrix is positive
definite. The symplectic spectrum of $\sigma$ is defined as follows: the
product $\Omega\sigma$ is similar to the antisymmetric matrix $\sigma
^{1/2}\Omega\sigma^{1/2}$ whose eigenvalues are of the type $\pm i\nu_{j}$,
$\nu_{j}>0$. The \emph{symplectic spectrum} of the covariance matrix is then
the sequence $(\nu_{1},...,\nu_{n})$; the numbers $\nu_{j}$ are called the
\emph{symplectic eigenvalues} of $\sigma$. An important property is that the
$\nu_{j}$ are symplectic invariants, i.e. they do not change under linear
symplectic changes of coordinates. This is because the eigenvalues of
$\Omega(S^{T}\sigma S)$ are the same\ as those of $\Omega\sigma$ since
$\Omega(S^{T}\sigma S)=S^{-1}(\Omega\sigma)S$ in view of the relation $S\Omega
S^{T}=\Omega$ characterizing linear symplectic transformations.

\section{Topological Form of the UP}

We have shown in \cite{FP,physreps,gostat} that condition (\ref{2}) is
equivalent to
\begin{equation}
c_{\mathrm{GR}}(W_{\sigma})\geq\pi\hbar=\tfrac{1}{2}h \label{4}%
\end{equation}
where $c_{\mathrm{GR}}(W_{\sigma})$ is the Gromov width (or: symplectic
capacity \cite{Birk,physreps}) of the covariance ellipsoid $W_{\sigma}%
:\frac{1}{2}z^{T}\sigma^{-1}z\leq1$. The argument goes as follows: the Gromov
width $c(W)$ of a subset $W$ of phase space $\mathbb{R}^{2n}$ is the supremum
of all numbers $\pi R^{2}$ such that the ball $B_{R}:|z|\leq R$ can be send
inside $W$ using symplectic transformations, linear or not (the notion is
related to Gromov's famous symplectic non-squeezing theorem \cite{Gromov} from
1985; see \cite{Birk,physreps} for a review). In view of Williamson's
diagonalization theorem \cite{williamson} there exists $S\in\operatorname*{Sp}%
(2n,\mathbb{R})$ such that
\[
S^{T}\sigma S=%
\begin{pmatrix}
\Lambda & 0\\
0 & \Lambda
\end{pmatrix}
\]
where $\Lambda$ is the diagonal matrix consisting of the symplectic
eigenvalues $\nu_{j}$, $j=1,...,n$, of the eigenvalues of the covariance
matrix. Symplectic capacities are symplectic invariants, hence it is
sufficient to assume that $\sigma=%
\begin{pmatrix}
\Lambda & 0\\
0 & \Lambda
\end{pmatrix}
$ which reduces the proof to the case where the ellipsoid $W_{\sigma}$ is
given by
\begin{equation}
\sum_{j=1}^{n}\frac{1}{2\nu_{j}}(x_{j}^{2}+p_{j}^{2})\leq1. \label{sign}%
\end{equation}
Assume now that we can squeeze the ball $B_{R}$ inside $W_{\sigma}$; this
requires that the projection of that ball onto each plane $x_{j},p_{j}$ of
conjugate coordinates has radius $R\leq\sqrt{2\nu_{j}}$, for each $j$ and
hence $R\leq\sqrt{2\nu_{\min}}$ where $\nu_{\min}=\inf\{\nu_{1},...,\nu_{n}%
\}$; it follows that $c(W_{\sigma})=\pi\nu_{\min}$. To prove the inequality
(\ref{4}) it is thus sufficient to show that $\nu_{\min}\geq\frac{1}{2}\hbar$.
We now exploit the condition (\ref{2}). Noting that the matrix $\sigma
+\tfrac{1}{2}i\hbar\Omega\geq0$ is similar to $+\tfrac{1}{2}i\hbar
\sigma^{-1/2}\Omega\sigma^{-1/2}$ this condition equivalent to
\[
I+\tfrac{1}{2}i\hbar\sigma^{-1/2}\Omega\sigma^{-1/2}\geq0.
\]
The characteristic polynomial of the matrix in the LHS is a product
\begin{equation}
P(t)=P_{1}(t)\cdot\cdot\cdot P_{n}(t) \label{pt}%
\end{equation}
of quadratic polynomials%
\begin{equation}
P_{j}(t)=t^{2}-2t+1-\frac{\hbar^{2}}{4\nu_{j}^{2}}; \label{ptj}%
\end{equation}
it follows that all the eigenvalues of that matrix are non-negative if and
only if $1\geq\hbar^{2}/4\nu_{j}^{2}$ for all $j=1,...,n$, that is is to
\begin{equation}
\nu_{j}\geq\frac{1}{2}\hbar. \label{nujh}%
\end{equation}

\section{The Symplectic Camel}

One way of stating Gromov's symplectic non-squeezing theorem is the following:
consider the phase space ball $B_{R}:|z|\leq R$; the orthogonal projection of
$B_{R}$ on any of the conjugate coordinates planes $x_{j},p_{j}$ is the circle
$x_{j}^{2}+p_{j}^{2}\leq R^{2}$, which has area $\pi R^{2}$. Suppose we
deform\ $B_{R}$ using a canonical transformation $f$; the set $f(B_{R})$ will
have same volume as $B_{R}$ (Liouville's theorem), but, in addition, the
orthogonal projection of $f(B_{R})$ on the planes $x_{j},p_{j}$ will always
have an area at least equal to $\pi R^{2}$ (but the area of the projection on
planes of non-conjugate variables can take arbitrarily small values). An
extension of the non-squeezing theorem has recently be proved by Abbondandolo
and Matveyev \cite{abmat}. It is well-understood only in the linear (and
affine) case, but this is sufficient for our purposes. Let $\mathbb{R}^{2k}$
($1\leq k\leq n$) be a symplectic subspace of $\mathbb{R}^{2n}$. This means
that the restriction of the symplectic product $\wedge$ to $\mathbb{R}^{2k}$
is non-degenerate, and hence itself a symplectic product. An elementary
example is the set of all coordinates $\{(x_{1},...,x_{k};p_{1},...,p_{k})\}$;
in fact every symplectic subspace can be obtained from the latter using linear
symplectic transformations. Let us denote by $B^{2k}(R)$ the ball with radius
$R$ centered at the origin in $\mathbb{R}^{2k}$; its volume is $(\pi
R^{2})^{k}/k!$. Let now $\Pi_{k}$ be the orthogonal projection of
$\mathbb{R}^{2n}$ onto $\mathbb{R}^{2k}$; obviously $\Pi_{k}B^{2n}%
(R)=B^{2k}(R)$. What Abbondandolo and Matveyev prove is that for very $S$ in
$\operatorname*{Sp}(2n,\mathbb{R})$ following inequality holds:%
\begin{equation}
\operatorname*{Vol}\Pi_{k}(SB^{2n}(R))\geq\operatorname*{Vol}\Pi_{k}%
(B^{2n}(R))=\frac{(\pi R^{2})^{k}}{k!}. \label{abbo}%
\end{equation}
When $k=n$ this inequality is trivial (it actually becomes an equality because
symplectic transformations are volume-preserving), and for $k=1$ it is a
reformulation of Gromov's non-squeezing theorem in the linear case. The
interesting point with this formula is that it provides us with a statement
valid for middle-dimensional symplectic spaces. Abbondandolo and Matveyev give
an example in \cite{abmat} showing that the inequality (\ref{abbo}) is not
always true for $k$ between $1$ and $n$ if we replace $S$ by an arbitrary
non-linear symplectic transformation; their counterexample is however of a
rather pathological nature, so one may conjecture that (\ref{abbo}) holds for
quite large classes of nonlinear symplectic transformations. This conjecture
has not yet been proved, and is an active area of research.

\section{Geometric Invariants: the Main Result}

The inequality (\ref{abbo}) allows us to prove the main result of this Letter:
assume that we project orthogonally the covariance ellipsoid $W_{\sigma}%
:\frac{1}{2}z^{T}\sigma^{-1}z\leq1$ on a $2k$-dimensional symplectic subspace
$\mathbb{F}_{2k}$ of $\mathbb{R}^{2n}$; the projection is an ellipsoid
$\Pi_{\mathbb{F}_{2k}}W_{\sigma}$ in $\mathbb{F}_{2k}$ whose volume
\begin{equation}
\operatorname*{Vol}(\Pi_{\mathbb{F}_{2k}}W_{\sigma})=\frac{(2\pi)^{k}}{k!}%
\nu_{1}\cdot\cdot\cdot\nu_{k} \label{pr}%
\end{equation}
satisfies the inequality%
\begin{equation}
\operatorname*{Vol}(\Pi_{\mathbb{F}_{2k}}W_{\sigma})\geq\frac{h^{k}}{2^{k}k!}
\label{pr1}%
\end{equation}

We first remark that it suffices to prove the inequality (\ref{pr1}) for
\textit{one} $2k$-dimensional symplectic subspace $\mathbb{F}_{2k}$ for it
will then hold for \emph{all }such subspaces. Let in fact $\mathbb{F}%
_{2k}^{\prime}$ be another such subspace with same dimension, and choose
symplectic bases $\mathcal{B}_{2k}=\{e_{1},...,e_{k};f_{1},...,f_{k}\}$ and
$\mathcal{B}_{2k}^{\prime}=\{e_{1}^{\prime},...,e_{k}^{\prime};f_{1}^{\prime
},...,f_{k}^{\prime}\}$ of $\mathbb{F}_{2k}$ and $\mathbb{F}_{2k}^{\prime}$,
respectively (i.e. $e_{i}\wedge e_{j}=f_{i}\wedge f_{j}=0$, $f_{i}\wedge
e_{j}=\delta_{ij}$ for $1\leq i,j\leq k$, and similar relations for the
elements of $\mathcal{B}_{k}^{\prime}$). Completing $\mathcal{B}_{2k}$ and
$\mathcal{B}_{2k}^{\prime}$ to full symplectic bases $\mathcal{B}_{2n}$ and
$\mathcal{B}_{2n}^{\prime}$ of $\mathbb{R}^{2n}$ (the \textquotedblleft
symplectic Gram-Schmidt theorem\textquotedblright, see e.g. \cite{Birk}, Ch.
1) the linear automorphism $S$ of $\mathbb{R}^{2n}$ defined by $S(e_{j}%
)=e_{j}^{\prime}$, $S(f_{j})=f_{j}^{\prime}$ for $1\leq j\leq n$ is
symplectic, and so is its restriction $S_{k}$ to the subspace $\mathbb{F}%
_{2k}$. Symplectic mappings being volume preserving, $S_{k}$ sends
$\Pi_{\mathbb{F}_{2k}}W_{\sigma}$ to an ellipsoid with same volume as
$\Pi_{\mathbb{F}_{2k}}W_{\sigma}$. Suppose now the covariance matrix diagonal
of the type $\lambda=%
\begin{pmatrix}
\Lambda & 0\\
0 & \Lambda
\end{pmatrix}
$; then, choosing $\mathbb{F}_{2k}=\mathbb{R}^{2k}$, identified with the set
of points $(x_{1},...,x_{k};p_{1},...,p_{k})$, we have, using formula
(\ref{sign}),
\[
\Pi_{\mathbb{R}^{2k}}W_{\lambda}:\sum_{j=1}^{k}\frac{1}{2\nu_{j}}(x_{j}%
^{2}+p_{j}^{2})\leq1.
\]
The volume of this projected ellipsoid is
\begin{equation}
\operatorname*{Vol}(\Pi_{\mathbb{R}^{2k}}W_{\lambda})=\frac{(2\pi)^{k}}{k!}%
\nu_{1}\cdot\cdot\cdot\nu_{k} \label{pr2}%
\end{equation}
and the inequality (\ref{pr1}) follows in this case since $\nu_{j}\geq\frac
{1}{2}\hbar$ for all indices $j$. The general case (\ref{pr1}) is easily
deduced. We first note that in view of the discussion of symplectic spaces
above, the inequality (\ref{pr2}) is preserved if we replace $\mathbb{R}^{2k}$
with an arbitrary symplectic subspace $\mathbb{F}_{2k}$. Next, writing
$\sigma=S^{T}\lambda S$ (Williamson's diagonal form \cite{williamson}) we have
$W_{\sigma}=S^{T}W_{\lambda}$; the claim now follows from the Abbondandolo and
Matveyev inequality (\ref{abbo}).

\section{Discussion}

In \cite{serafini1,serafini2} Serafini studies the \textquotedblleft universal
symplectic invariants\textquotedblright\ introduced by Dodonov \cite{dodo}.
Working in units in which $\hbar=2$, he denotes by $\Delta_{j}^{n}$ the
principal minor of order $2j$ of $\Omega\sigma$,\textit{ i.e.}, the sum of the
determinants of all the principal submatrices of order $2j$ of $\Omega\sigma$
(by convention $\Delta_{0}^{n}=1$). It is easy to check that the numbers
$\Delta_{j}^{n}$ are just the coefficients of the characteristic polynomial of
the matrix $\Omega\sigma$, and hence
\begin{equation}
\Delta_{j}^{n}=\sum_{C(n,j)}\left(  \prod\limits_{k\in C(n,j)}\nu_{k}%
^{2}\right)  \label{delta}%
\end{equation}
where the sum runs over all the possible combinations $C(n,j)$ of $j$ integers
$\leq n$. A rapid comparison of formulas (\ref{pr2}) and (\ref{delta}) shows
that the numbers $\Delta_{k}^{n}$ are related to the volumes of the
projections of the covariance ellipsoid by the formula%
\[
\Delta_{j}^{n}=\left(  \frac{j^{!}}{(2\pi)^{j}}\right)  ^{2}\sum
_{\mathbb{F}_{2j}\in\operatorname*{Sp}_{2j}}\left(  \Pi_{\mathbb{F}_{2j}%
}W_{\sigma}\right)  ^{2}%
\]
where $\operatorname*{Sp}_{2k}$ is the set of all $n!/j!(n-j)!$ symplectic
coordinate subspaces of $\mathbb{R}^{2n}$.

The terms $\left(  \Pi_{\mathbb{F}_{2j}}W_{\sigma}\right)  ^{2}$ which are of
a geometric nature contain more information that the $\Delta_{j}^{n}$ which
are defined in algebraic terms, and should thus given a privileged status when
studying invariants of continuous variables quantum systems.

\end{document}